\documentclass{article}
\usepackage[utf8]{inputenc}
\usepackage{graphicx}
\usepackage{authblk}

\usepackage{placeins}
\makeatother

\usepackage{listings}
\usepackage{booktabs}
\usepackage{fancyhdr}
\fancypagestyle{title}{
  \fancyhf{}
  \fancyfoot[C]{\thepage}
  \fancyhead[R]{DESY 17-096}
}%

\usepackage{color}

\usepackage{tikz}
\usetikzlibrary{shapes,arrows.meta}
\tikzset{%
  >={Latex[width=2mm,length=2mm]},
            base/.style = {rectangle, rounded corners, draw=black,
                           minimum width=4cm, minimum height=1cm,
                           text centered, font=\sffamily},
     bluenode/.style = {base, fill=blue!30},
     rednode/.style = {base, fill=red!30},
     greennode/.style = {base, fill=green!30},
     orangenode/.style = {base, minimum width=2.5cm, fill=orange!15,
                           font=\ttfamily},
                       }
\tikzstyle{decision} = [diamond, draw, fill=blue!20,
    text width=6.5em, text badly centered, node distance=3cm, inner sep=0pt]

\title{Enabling rootless Linux Containers in multi-user environments: the {\sl udocker} tool}

\author[1]{\small{Jorge Gomes}}
\author[2]{Emanuele Bagnaschi}
\author[3]{Isabel Campos}
\author[1]{\\ Mario David}
\author[1]{Lu\'is Alves}
\author[1]{Jo\~ao Martins}
\author[1]{Jo\~ao Pina}
\author[3]{Alvaro L\'opez-Garc\'{i}a}
\author[3]{Pablo Orviz}

{\small {
\affil[1]{Laborat\'orio de Instrumenta\c{c}\~ao e F\'{i}sica Experimental
de Part\'{i}culas (LIP), Lisboa, Portugal}
\affil[2]{Deutsches Elektronen-Synchrotron (DESY), 22607 Hamburg, Germany}
\affil[3]{IFCA, Consejo Superior de Investigaciones Cient\'{i}ficas-CSIC, Santander, Spain}

}}

\begin{document}

\maketitle
\thispagestyle{title}

\begin{abstract}
Containers are increasingly used as means to distribute and run Linux services and applications. In this paper we describe the architectural design and implementation of \emph{udocker}, a tool which enables the user to execute Linux containers in user mode. We also present a few practical applications, using a range of scientific codes characterized by different requirements: from single core execution to MPI parallel execution and execution on GPGPUs.
\end{abstract}

\section{Introduction}

Technologies based on Linux containers have become very popular among software developers and system administrators. The main reason behind this success is the flexibility and efficiency that containers offer when it comes to packing, deploying and running software.

A given software can be containerized together with all its dependencies in such a way that it can be seamlessly executed regardless of the Linux distribution used by the designated host systems.
This is achieved by using advanced features of modern Linux kernels \cite{KERNEL}, namely {\sl control groups} and {\sl namespaces} isolation \cite{CG,NAMESP}. Using both features, a set of processes can be placed in a fully isolated environment (using {\sl namespaces isolation}), with a given amount of resources, such as CPU or RAM, allocated to it (using {\sl control groups}). We call ``container'' this encapsulated group of processes.

Containers do not need to boot an operating system kernel, instead they share the one of the host system.
Therefore they can be seen as a lightweight approach to virtualization, when compared with conventional Virtual Machines (VM) such as the ones provided using para-virtualization or hardware virtual machines.

As a consequence, a given physical system may host many more containers than virtual machines, as the first ones use less hardware resources (notably memory). Moreover, transferring, starting and shutting down containers is usually much faster than performing the same operations on a conventional virtual machine.

The idea of operating-system-level virtualization is not new. The origin can be traced back to the {\sl chroot} system call (introduced in 1979), which enables changing the root directory for a given process and its children. The concept was further extended with the introduction of the jailed systems of BSD \cite{JAILBSD} (released in 1998). However the developments required at the level of the Linux kernel to have a consistent and robust implementation have taken a few years more to crystallize. The Control groups technology, also known as cgroups, is present in the Linux kernels since version 2.6.24 (released in 2008). Support for namespaces was first introduced in the Linux kernel 2.4.19 (released in 2002), with additional namespaces and enhancements being added since. However it was only with the introduction of Docker~\cite{DOCKER} that containers gained widespread adoption in Linux.

Docker simplifies the creation, distribution and execution of Linux containers. Besides providing the tools to build containers and execute them, it also supports container repositories (such as the Dockerhub \cite{HUB}), enabling the upload and the download of container images.

Docker container images make use of a layered filesystem where changes are added as new layers. Using this approach, a new Docker image can be created from an existing one simply by reusing and sharing its layers. Sharing layers among images saves storage space and minimizes downloads.

Because of these very appealing features described above, Docker is now widely used to containerize both services and applications. However Docker presents limitations especially when it comes to deploy containers in multi-user systems. In Docker, processes within the container are normally executed with root privileges under the Docker daemon process tree, thus escaping to resource usage policies, accounting controls, and process controls that are imposed to normal users. In addition, users authorized to access the Docker remote API can easily gain privileged access to the host system. Due to these limitations Docker is not generally available in multi-user systems.

Unfortunately most scientific computing resources such as High Performance Computing (HPC) systems, Linux clusters or grid infrastructures are multi-user systems. Therefore the adoption of Docker in these infrastructures has been very restricted.

Another limitation of Docker concerns the support for parallel applications that require network communication across multiple hosts. For security reasons Docker uses network namespaces, making direct communication between containers across different hosts more difficult with respect to normal processes. The use of network namespaces can be disabled but, opting for this solution, results in a system security issue, as privileged processes running within the container may interfere with the host network stack. Access to host devices from within the container is similarly problematic for the same reasons.

The Docker layered filesystem has many advantages but it is also slower than accessing the host filesystem directly. Furthermore files in existing layers cannot be deleted or changed, instead they are hidden or overridden by an upper layer. Consequently container images grow in size when modified and the only way to keep their size at acceptable level is to rebuild them from scratch.

The applicability of Docker to scientific computing has been hampered by the limitations we have highlighted in the prevision paragraphs. As a consequence, alternative solutions such as Singularity \cite{SINGULARITY} or Shifter \cite{SHIFTER} have been developed. However these solutions also have their own limitations. They generally require administrator privileges both for installation and execution, or alternatively require recent Linux kernel features such as unprivileged namespaces which are still not widely available in scientific computing systems. 

In this work we address the problematic of executing Docker containers in user space, i.e. without installing additional system software, without requiring any administrative privileges and in a way that respects resource usage policies, accounting and process controls. Our aim is to empower users to execute applications encapsulated in Docker containers easily in any Linux system including computing clusters without system administrator intervention and regardless of Docker being locally available.

In section 2 we will describe the architectural elements of the tool we have developed to address the above mentioned problems and shortfalls. The remaining of the paper is devoted to describe the practical applicability of this middleware. We describe how our solution can be applied to three very different use cases which we believe are representative of the current scientific-computing landscape.

The first use case describes an application highly complex due to the interplay of library dependencies, legacy code, library and software requirements. The second case addresses the usage of containers in MPI parallel applications using Infiniband and/or TCP/IP. Finally we also show how to target specialized hardware accelerators such as GPGPUs.

All the tools presented here are open-source and may be downloaded from our public repository \cite{UDOCKER}.

\section{Architectural design}

To overcome the limitations mentioned in section 1 we have explored the possibilities available to run applications encapsulated in Docker containers in user space in a portable way. Our technical analysis is as follows.

\subsection{Technical Analysis}

Most scientific applications do not require the full set of Docker features. These applications are usually developed to be executed in multi-user shared environments by unprivileged users. Therefore features such as isolation are not strictly needed as long as the applications are executed without privileges. The fundamental features that make Docker appealing as a means to execute scientific applications are thus the following:

\begin{itemize}
\item the ability to provide a separate directory tree where the application with all its dependencies can execute independently from the target host;
\item possibility of mounting host directories inside the container directory tree, which is convenient for data access;
\item easy software packaging with Dockerfiles;
\item reuse of previously created images;
\item easy sharing and distribution of images via Docker repositories such as Docker Hub and many others;
\item simple command line and REST interfaces.
\end{itemize}

Several of these features such as creating containers and pushing them into repositories can be accomplished from the user desktop or portable computer using Docker directly. Therefore we assume that Docker can be used to prepare the containers and that the focus should rather be put on enabling the execution of the created containers in systems where Docker is unavailable.

The most complex aspect is to provide a chroot-like functionality so that Docker containers can be executed without conflicting with the host operating system environment. The chroot system call requires privileges and therefore would collide with our objective of having a tool that would not require privileges or installation by the system administrator. Three alternative approaches to implement chroot-like functionality without privileges were identified:

\begin{itemize}
\item use unprivileged {\tt User Namespaces};
\item use PTRACE to intercept calls that handle pathnames;
\item use LD\_PRELOAD to intercept calls that handle pathnames.
\end{itemize}

The use of unprivileged {\tt User Namespaces} allows a non-privileged user to take advantage of the Linux namespaces mechanism. This approach is only usable since kernel 3.19. Unfortunately the implementation of unprivileged {\tt User Namespaces} exposes the processes running inside the container to limitations especially in terms of mappings of group identifiers. Furthermore this approach does not work on certain distributions such as CentOS 6 and CentOS 7, which do not provide kernels having the necessary features. Due to these limitations this approach was initially discarded.

The PTRACE mechanism enables tracing of system calls making possible to change their calling parameters in run time.
System calls that use pathnames can be traced and dynamically changed so that references to pathnames within the container can be dynamically expanded into host pathnames prior to the system calls and upon their return.
The biggest drawback of using PTRACE to implement chroot-like functionality is the impact on performance. An external process needs to trace the execution of the application, stop it before each system call, change its parameters when they reference pathnames, and continue from the same point. If the system call returns pathnames a second stop must be performed when the call returns to perform the required changes. In older kernels (such as in CentOS 6) all system calls need to be traced. In more recent kernels is possible to use PTRACE with SECCOMP filtering to perform selective system call tracing thus allowing to restrict interception only to the set of calls that manipulate pathnames. The impact of using PTRACE depends on the availability of SECCOMP and on how frequently the application invokes system calls.

The LD\_PRELOAD mechanism allows overriding of shared libraries. In practice this approach allows to call wrapping functions where the pathnames can be changed before and after invoking the actual system functions. The LD\_PRELOAD mechanism is implemented by the dynamic loader {\it ld.so}. The dynamic loader is an executable that finds and loads shared libraries and prepares programs for execution. The dynamic loader also provides a set of calls that enable applications to load further dynamic libraries in run time. When a shared library pathname does not contain a slash the dynamic loader searches for the library in the directory locations provided by:

\begin{itemize}
\item {\tt DT\_RPATH} dynamic section attribute of the ELF executable;
\item {\tt LD\_LIBRARY\_PATH} environment variable;
\item {\tt DT\_RUNPATH} dynamic section attribute of the ELF executable;
\item cache file {\tt /etc/ld.so.cache};
\item default paths such as {\tt /lib64}, {\tt /usr/lib64}, {\tt /lib}, {\tt /usr/lib}.
\end{itemize}

Due to this behavior, libraries from the host system may end-up being loaded instead of the libraries within the chroot-environment. Furthermore the following limitations apply:

\begin{itemize}
\item this approach depends on dynamic linking and does not work with statically linked executables;
\item the absolute pathname to the dynamic loader is encoded in the executables, leading to the invocation of the host dynamic loader instead of the dynamic loader provided in the container;
\item dynamic executables and shared libraries may also reference other shared libraries, if they have absolute pathnames then they may be loaded from locations outside of the container.
 \end{itemize}

\subsection{The {\sl udocker} tool}

To validate the concept we developed a tool called {\sl udocker} that combines the pulling, extraction and execution of Docker containers without privileges. {\sl udocker} is an integration tool that incorporates several execution methods giving the user the best possible options to execute their containers according to the target host capabilities. {\sl udocker} is written in Python and aims to be portable. {\sl udocker} has eight main blocks:

\begin{itemize}
\item user command line interface;
\item self installation;
\item interface with Docker Hub repositories;
\item local repository of images;
\item creation of containers from images;
\item local repository of containers;
\item containers execution;
\item specific execution methods.
\end{itemize}

{\sl udocker} provides a command line interface similar to Docker and provides a subset of its commands aimed at searching, pulling and executing containers in a Docker like manner.

The self installation allows a user to transfer the {\sl udocker} Python script and upon the first invocation pull any additional required tools and libraries which are then stored in the user directory. This allows {\sl udocker} to be easily deployed and upgraded by the user himself. {\sl udocker} can also be installed from a previously downloaded tarball.

The Docker images are composed of filesystem layers. Each layer has metadata and a directory tree that must be stacked by the layered filesystem to obtain the complete image. The Docker layered filesystem is based on UnionFS \cite{UNIONFS}. In UnionFS file deletion is implemented by signaling the hiding of files in the lower layers via additional files that act as markers {\it white-outs}.
We therefore implemented the pulling of images by downloading the corresponding layers and metadata using the Docker Hub REST API. A simple repository has been implemented where layers and metadata are stored. Layers are shared across the images to save space. The repository is by default placed in the user home directory.
To prepare a container directory tree, the several layers are sequentially extracted over the previous ones respecting the {\it white-outs}. File protections are then adjusted so that the user can access all container files and directories. The container directory tree is also stored in the local repository. The container execution is achieved with a chroot-like operation using the previously prepared container directory tree.

{\sl udocker} implements the parsing of Docker container metadata and supports a subset of metadata options namely: the command to be invoked when the container is started, mount of host directories, setting of environment variables, and the initial working directory.


The PTRACE execution method was implemented using PRoot \cite{PROOT}, a tool that provides chroot-like functionality using PTRACE. PRoot also provides mapping of host files and directories into the container enabling access from the container to host locations. We had to develop fixes for PROOT to make SECCOMP filtering work together with PTRACE due to changes introduced in the Linux kernel 3.8 and above. A fix for proper clean-up of temporary files was also added.

The LD\_PRELOAD method is based on Fakechroot \cite{FAKECHROOT} a library that provides chroot-like functionality using this mechanism, and is commonly used in {\it Debian} to install a base system in a subdirectory. Fakechroot provides most of the wrapping functions that are needed. However due to the previously described behavior of the dynamic loader, applications running under Fakechroot may unwillingly load shared libraries from the host. {\sl udocker} implements several strategies to address these shortcomings.

\begin{itemize}
\item Replacement of the dynamic loader pathname in the executables header, thus forcing the execution of the loader provided with the container.
\item Replacement of DT\_RPATH and DT\_RUNPATH in the executables, forcing the search paths to container locations.
\item Replacement of shared library names within the headers of executables and libraries, forcing them to container locations.
\item Extraction of shared libraries pathnames from ld.so.cache which are then modified and added to LD\_LIBRARY\_PATH.
\item Interception of changes to LD\_LIBRARY\_PATH forcing the paths to container locations.
\item Prevent the dynamic loader from accessing the host ld.so.cache
\item Prevent the dynamic loader from loading libraries from the host {\tt /lib64}, {\tt /usr/lib64}, {\tt /lib}, {\tt /usr/lib}.
\end{itemize}

Changes to executables and libraries are performed using {\sl PatchELF} \cite{PATCHELF}. {\sl PatchELF} was enhanced to manage the pathname prefixes of: the dynamic loader within executables, library dependencies and their paths in executables and shared libraries. The {\sl ld.so} executable within the container can also be modified by {\sl udocker} to prevent the loading of shared libraries from the host. All the above changes are performed automatically by {\sl udocker} depending on the execution mode selected. The Fakechroot library was heavily modified to address multiple limitations, change executables and libraries in run time, and to provide better mapping of host directories and files inside the container.

The support for unprivileged {\tt User Namespaces} and rootless containers was implemented using runC \cite{RUNC}, a tool for spawning and running containers according to the Open Containers Initiative OCI \cite{OCI} specification. {\sl udocker} translates Docker metadata and the command line arguments to build a matching OCI configuration spec that enables the container execution in unprivileged mode.

All external tools including PRoot, PatchELF, runC and Fakechroot are provided together with {\sl udocker}. The binary executables are statically compiled to be used across multiple Linux distributions unchanged and increasing portability. {\sl udocker} itself has a minimal set of Python dependencies and can be executed in a wide range of Linux systems. {\sl udocker} can be either deployed and managed entirely by the end-user or centrally deployed by a system administrator. The {\sl udocker} execution takes place under the regular userid without requiring any additional privilege.

\section{Description of Basic Capabilities}

In this section we provide a description of the main capabilities of {\sl udocker}. As previously stated, it mimics a subset of the Docker syntax to facilitate its adoption and use by those already familiar with Docker tools. As pointed out in sections 1 and 2, it does not make use of Docker nor requires its presence.
The current implementation is limited to the pulling and execution of Docker containers. The actual containers should be built using Docker and Dockerfiles. {\sl udocker} does not provide all the Docker features since it's not intended as a Docker replacement but oriented to providing a run-time environment for containers execution in user space.
The following is a list of examples, see \cite{UDOCKER} for a complete list.

\begin{itemize}
\item Pulling containers from Docker Hub or from private Docker repositories.

\begin{verbatim}
udocker pull docker.io/repo_name/container_name
\end{verbatim}

The image of the container with its layers is downloaded to the directory specified by the user. The location of that directory is controlled by an environment variable {\tt \$UDOCKER\_DIR} that can be defined by the user. It should point to a directory in a filesystem where there is enough capacity to unroll the image. The default location is {\tt \$HOME/.udocker}.

\item  Once the image is downloaded, the container directory tree can be obtained from the container layers using the option {\it create}.

Upon exit {\sl udocker} displays the identifier of the newly created container, a more understandable name can be associated to the container identifier using the option {\it name}. In the example below, the content of the {\tt ROOT} directory of a given container is shown. The {\tt ROOT} is a subdirectory below the container directory.

\begin{verbatim}
$shell> udocker create docker.io/repo_name/container_name
95c22b84-1868-332b-9bf0-2e056beafb00

$shell> udocker name 95c22b84-1868-332b-9bf0-2e056beafb00 \
        my_container

$shell> ls $HOME/.udocker
bin  containers  layers  lib  repos

$shell> ls .udocker/containers/my_container/ROOT/
bin   dev  home  lib64       media  opt   root  sbin  sys  usr
boot  etc  lib   lost+found  mnt    proc  run   srv   tmp  var
\end{verbatim}

\item Install software inside the container.

\begin{verbatim}
udocker run  --user=root container_name  \
        yum install -y firefox pulseaudio gnash-plugin
\end{verbatim}

In this example we used the capability to partially emulate root in order to use {\tt yum} to install software packages in the container. No root privileges are involved in this operation. Other tools that make use of user namespaces are not capable of installing software in this way.

\item Share hosts directories, files and devices with the container.

Execute making host directories visible inside the container.

\begin{verbatim}
udocker run -v /var -v /tmp -v /home/x/user:/mnt \
            container_name  /bin/bash
\end{verbatim}

In this example we executed a container with the host directories {\tt /var} and {\tt /tmp} visible inside the container.
The host directory {\tt /home/x/user} is also mapped into the container directory {\tt /mnt}. Existing files in the container directories acting as mount points will be obfuscated by the content of the host directories.

\item Accessing the network as a host application. Contrary to the default Docker behavior {\sl udocker} does not deploy an additional network environment, and uses the host network environment unchanged.

\begin{verbatim}
udocker run container_name /usr/sbin/ifconfig
\end{verbatim}

\item Run graphics natively accessing the desktop X-Windows server directly.

\begin{verbatim}
udocker run --bindhome --hostauth --hostenv \
        -v /sys -v /proc -v /var/run -v /dev --user=jorge \
        --dri container_name  /usr/bin/xeyes
\end{verbatim}

Since {\sl udocker} is running inside the user environment it can transparently access devices and functionality available to the parent user process such as the X Windows System.

\end{itemize}

\subsection{Execution modes}
The execution modes described in table~\ref{table:exec_modes} correspond to the different approaches implemented using PRoot, Fakechroot, runC and Singularity. All modes are interchangeable, the execution mode of a given container can be changed as needed.

The {\tt P} modes are the most interoperable, they support older and newer Linux kernels.

The {\tt F} modes can be faster than {\tt P} if the application makes heavy use of system calls. The {\tt F} modes require libraries that are provided with {\sl udocker} for specific distributions such as CentOS 6 and 7, Ubuntu 14 and 16 among others.

The {\tt R} mode requires a recent Linux kernel and currently only offers access to a very limited set of host devices which unfortunately prevents its usage with GPGPUs and low latency interconnects.

The {\tt S} mode uses Singularity if locally installed.

\begin{table}[h!]
  \begin{center}
    \begin{tabular}{l|c|l} \toprule
    \textbf{Mode} & Engine & Characteristics \\
    \hline
    \textbf{P1} & PRoot & uses SECCOMP filtering (default mode)\\
    \textbf{P2} & PRoot & without SECCOMP filtering\\
    \textbf{F1} & Fakechroot & uses loader as 1st argument and LD\_LIBRARY\_PATH\\
    \textbf{F2} & Fakechroot & uses modified loader to prevent loading from host\\
    \textbf{F3} & Fakechroot & ELF headers of executables and libraries are modified\\
    \textbf{F4} & Fakechroot & ELF headers are modified dynamically if needed\\
    \textbf{R1} & runC & uses namespaces\\
    \textbf{S1} & Singularity & if locally installed, executes in sandbox mode\\
    \end{tabular}
  \end{center}
  \caption{Execution modes.}
  \label{table:exec_modes}
\end{table}

The selection of a given execution mode is accomplished with the {\tt setup} command as in the following example.

\begin{verbatim}
udocker setup --execmode=F3 container_name
\end{verbatim}

\subsection{Security}

Since root privileges are not involved, any operation that requires such privileges will not be possible. The following are examples of operations that are not possible:

\begin{itemize}
\item Accessing host protected devices and files.
\item Listening on $TCP/IP$ privileged ports (range below 1024).
\item Mount file-systems.
\item Change userid or groupid.
\item Change the system time.
\item Change routing tables, firewall rules, or network interfaces.
\end{itemize}

The {\tt P} and {\tt F} modes do not provide isolation features such as the ones offered by Docker. If the containers content is not trusted then they should not be executed with these modes, as they will be executed inside the user environment and can easily access host files and devices. The mode {\tt R1} provides better isolation by using namespaces.

The containers data is unpacked and stored in the user home directory or other location of choice. Therefore the containers data will be subjected to the same filesystem protections as other files owned by the user. If the containers have sensitive information the files and directories should be adequately protected by the user.

In the following sections we describe three advanced scientific applications, widely used by the respective communities, in which the applicability of the developed solution is demonstrated.

\section{Complex library dependencies: MasterCode}

The current pattern of research activity in particle physics requires in many cases the use of complex computational frameworks, often characterized by being composed of different programs and libraries developed by independent groups.

Moreover, in the past twenty years the average complexity of the standalone computational tools, which are part of these frameworks, has increased significantly and nowadays they usually depend on a complex network of external libraries and dependencies.

At the same time, the computational resources required by physics studies have noticeably increased in comparison with the past and the use of computer clusters composed of hundreds, if not thousands, of nodes is now common. To address this issue, scientific collaborations split the computational load across different sites, whose operating system and software environment usually differs.

As such, we investigate how the use of {\sl udocker} could ease the deployment of complex scientific applications, allowing for an easy and fast setup across heterogeneous systems. As an added value, we also observe that the use of exactly the same execution environment, where the application has supposedly been validated, protects against possible unexpected issues when moving between different sites with different compilers and libraries.

A prime example of modern complex applications is given by the frameworks used to perform global studies of particle physics models.
The role of these applications, is to compare the theoretical predictions from hypothetical models with the observations performed by experiments,
through probabilistic studies (either frequentist or Bayesian), of the parameters that enter in the definition of these new models.

All should be done by thoroughly covering as many different measurements and observations as possible. Since new physics can manifest itself with different effects according to the specific experiment (or observation), it is not consistent to perform independent studies for each one separately.

On the computational side, this translates in the practical requirement of interfacing a plethora of different codes, each usually developed with the aim of computing the theoretical prediction of at most a few observables, with a framework that performs the comparison with experimental data and that manages the sampling of the parameter space.

Several groups in the particle physics community have developed their own independent frameworks~\cite{ref:MasterCode, ref:GAMBIT, ref:Fittino, ref:SuperBayes}. As a concrete example, for our studies, we consider {\tt MasterCode}~\cite{ref:MasterCode}.

\subsection{Code structure}

As explained in the previous section, {\tt MasterCode} depends on a wide range of external tools and libraries,
which are tightly integrated into the main code-base. The full list of direct dependencies is the following:

\begin{itemize}
    \item GNU autotools and GNU {\tt make}.
    \item GNU toolchain (C/C++, Fortran77/90 compiler, linker, assembler).
    \item {\tt wget}, {\tt tar}, {\tt find}, {\tt patch}.
    \item {\tt Python} 2.7 or {\tt Python}~$\geq$~3.3.
    \item {\tt Cython~$\geq$~0.23}~\cite{ref:cython}.
    \item {\tt ROOT $\geq$ 5}~\cite{ref:ROOT} and its dependencies.
    \item {\tt numpy and its dependencies}~\cite{ref:numpy}.
    \item {\tt Matplotlib}~\cite{ref:matplotlib} and its dependencies.
    \item {\tt SQLite3}.
    \item {\tt SLHAlib}~\cite{ref:slhalib}.
    \item {\tt MultiNest}~\cite{ref:multinest}.
\end{itemize}

Moreover, it includes the following codes and their dependencies:

\begin{itemize}
    \item {\tt SoftSUSY}~\cite{ref:softsusy}.
    \item {\tt FeynWZ}~\cite{ref:feynwz}.
    \item {\tt FeynHiggs}~\cite{ref:feynhiggs}, {\tt HiggsBounds}~\cite{ref:higgsbounds} and {\tt HiggsSignals}~\cite{ref:higgssignals}.
    \item {\tt micrOMEGAs}~\cite{ref:micromegas}.
    \item {\tt SSDARD}~\cite{ref:ssard}.
    \item {\tt SDECAY}~\cite{ref:sdecay}.
    \item {\tt SuFla}~\cite{ref:sufla}.
\end{itemize}

The GNU toolchain is required not only at compile-time but at run-time as well. Some of the applications used to compute the theoretical predictions are meta--codes that generate new code to be compiled and executed according to the parameter space point under analysis.

The capability of {\sl udocker} in easing the use of {\tt MasterCode} was tested, by building a {\tt Docker} image based on {\tt Fedora} 23, with all required libraries and tools installed. After, the image was deployed successfully at three different sites, where the physics analyses are usually run. We also successfully interfaced the execution of {\sl udocker} with the local batch-systems.

To evaluate the performance impact of using {\sl udocker}, we compared the compilation time of {\tt MasterCode} with different setups, on a local workstation (see table~\ref{table:mastercodemachine} for the specifications).
\begin{table}
\parbox{.45\linewidth}{
  \begin{center}
    \begin{tabular}{@{}l||lc@{}} \toprule
      \textbf{OS} & & Fedora 23  \\
      \textbf{CPU} & & Intel Core i5 650   \\
      \textbf{RAM} & & 8 GB   \\
      \textbf{File System} & & ext4  \\
      \bottomrule
    \end{tabular}
  \end{center}
  \caption{Test machine specifications.}
  \vspace{0.85cm}
  \label{table:mastercodemachine}
}
\parbox{.45\linewidth}{
  \begin{center}
    \begin{tabular}{@{}l||lc@{}} \toprule
      \textbf{gcc} & & Red Hat 5.3.1-6  \\
      \textbf{Python} & & 3.4.3   \\
      \textbf{Cython} & & 0.23.4  \\
      \textbf{ROOT} & & 5.34/36  \\
      \bottomrule
    \end{tabular}
  \end{center}
  \caption{Software version of the most important tools and libraries used in our {\tt MasterCode} benchmarks.}
  \label{table:software}
}

\end{table}

\subsection{Execution flow}

\begin{figure}[t]
  \centering
\begin{tikzpicture}[node distance=1.5cm,
    every node/.style={fill=white, font=\sffamily}, align=center]
  \node (samplstart)   [bluenode]                                 {Sampling start};
  \node (multisel)     [orangenode, below of=samplstart]          {{\tt MultiNest} selects a point};
  \node (runcodes)     [orangenode, below of=multisel]            {Run computer codes to \\ compute physical observables};
  \node (chi2)         [orangenode, below of=runcodes]            {Compute $\chi^2$};
  \node (storage)      [greennode, below of=chi2]                 {Save data to storage \\using {\tt SQLite}};
  \node (convergence)  [decision, below of=storage, yshift=0.6cm] {convergence?};
  \node at (-1.8,-8) {No};
  \node at (0.6,-10) {Yes};
  \coordinate[left of=storage, xshift=-4cm] (fakenode);
  \node (samplend)     [rednode, below of=convergence,yshift=-1cm]  {Sampling end};
  \draw[->]     (samplstart)  -- (multisel);
  \draw[->]     (multisel)    -- (runcodes);
  \draw[->]     (runcodes)    -- (chi2);
  \draw[->]     (chi2)        -- (storage);
  \draw[->]     (storage)     -- (convergence);
  \draw[->]     (convergence) -- (samplend);
  \draw[-]      (convergence) -| (fakenode);
  \draw[->]     (fakenode)    |- (multisel);
\end{tikzpicture}
\caption{{\tt MasterCode} execution flow during the sampling phase.}
\label{fig:mastercodeflow}
\end{figure}
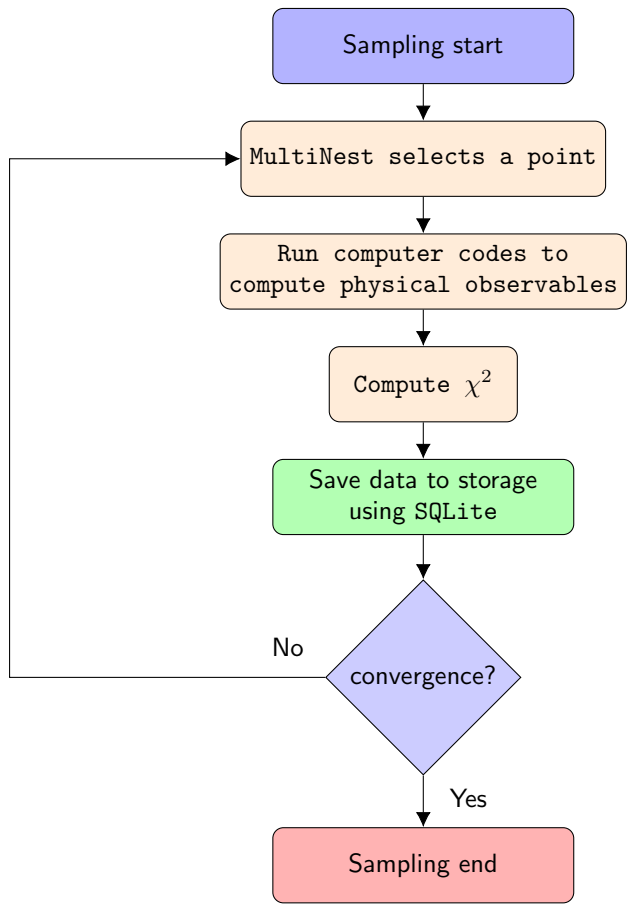

The execution of a full analysis in {\tt MasterCode} consists of three different different stages.

The first -- and most complex -- is the sampling of the parameter space of the model. In this phase of the execution,
the {\tt MultiNest} algorithm is used to sample the parameter space of the model. For each point in the parameter space,
a call is made to each of the external codes used to compute the theoretical predictions. The predictions are then compared with
the experimental constraints and the likelihood of the point under analysis is computed.
All the computed information is stored on the disk in an SQLite database for further analysis.

The main executable consists of a Python script that loads, through the {\tt ctypes} interface, the {\tt MultiNest} library,
while Cython interfaces are used to call the theory codes. Cython is also used to load an internal library, written in C++,
that computes the likelihood (i.e.~the $\chi^2$) of the point under scrutiny. Important computational, I/O and storage resources are required
for a fast execution of the sampling.

In practice, the parameter space of new physics models
is usually too large to be sampled by a single instance of {\tt MasterCode}.
To overcome this obstacle, it is usually segmented in sub-spaces, for each one of which a separate sampling campaign is run.
The execution flow is schematically depicted in fig.~\ref{fig:mastercodeflow}.

The second phase proceeds after the end of the sampling campaigns. The results of the various parameter--space sub--segments,
stored in a large number of {\tt SQLite} databases, are merged through the use of dedicated Python scripts.
To reduce the total size of the dataset and ease the analysis phase a selection filter is applied to the sampled points.
This step involves only I/O facilities and very often the performance limiting factor is the underlying file system.

The last stage is the physics analysis. In this step, one and two--dimensional {\tt ROOT} histograms, defined for various
physics observables and/or parameters, are filled with the likelihood information. From the computational perspective,
it involves reading the databases produced in the previous step, whose size is commonly of the order of 10--50 GB (but can reach the terabyte for complex models),
plus some lightweight computations to update the likelihood if needed.

\subsection{Benchmarking {\sl udocker} with MasterCode}

\begin{figure}
  \centering
  \includegraphics[width=0.92\textwidth]{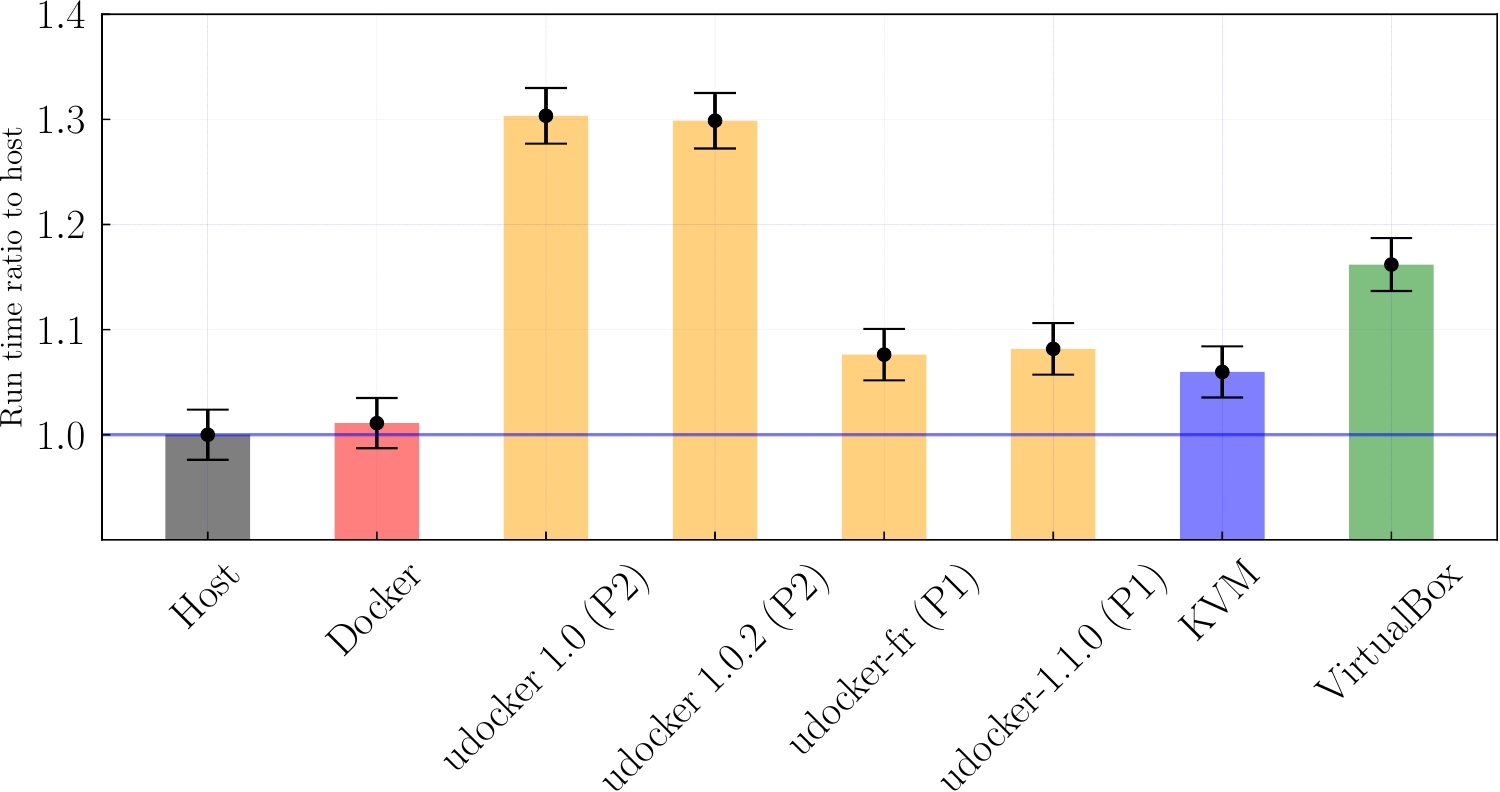}
  \caption{Compilation time for the {\tt MasterCode} framework using the different setups we considered in our benchmark, averaged over ten repetitions of the benchmark and then
    shown as a ratio over the result obtained on the host. The black bars show the standard mean deviation. The color coding is the following:
    in gray we show the result obtained on the host machine. In orange we show the compilation time using different {\sl udocker} setups.
    Red is used to show the {\sl Docker} result, while blue and green are used to plot the performances of {\sl KVM} and {\sl VirtualBox} respectively.}
  \label{fig:mastercodecompile}
\end{figure}

\begin{figure}
  \centering
  \includegraphics[width=\textwidth]{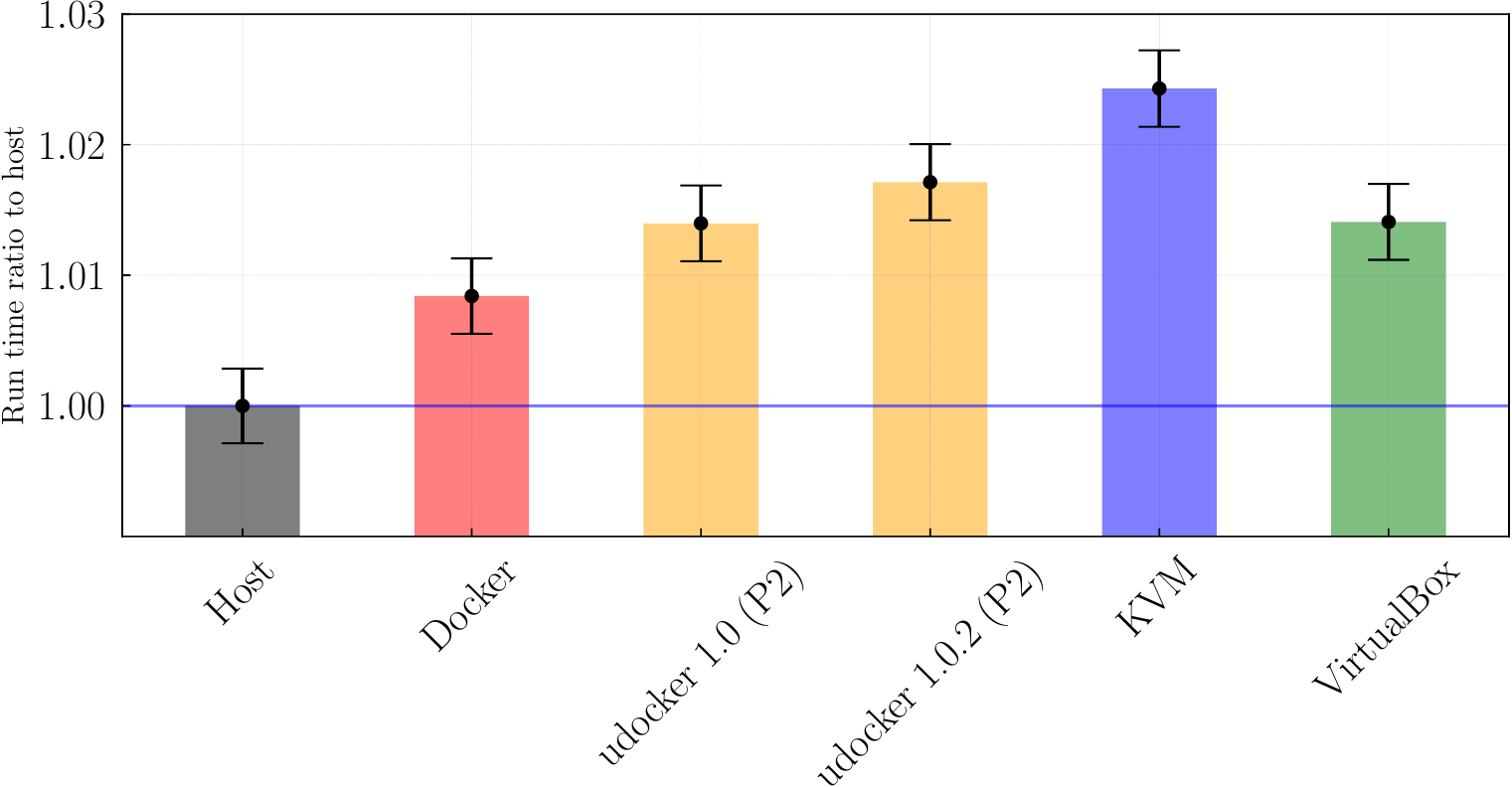}
  \caption{Running time for a test {\tt MasterCode} CMSSM sampling campaign, on our test machine, using different setups. Line colors and styling as in Figure~\ref{fig:mastercodecompile}.}
  \label{fig:mastercodecmssm}
\end{figure}

To verify if the use of {\sl udocker} has any significant performance impact on {\tt MasterCode}, two benchmarks where performed in different environments:

\begin{itemize}
\item The compilation time of {\tt MasterCode}.
\item Comparison for a fixed pseudo-random number sequence, the running time for a restricted sampling of the so-called Constraint Minimal Supersymmetric Standard Model (CMSSM).
\end{itemize}

The environments that we considered for our comparison are: the bare host, three different versions of {\sl udocker}, {\sl Docker}, {\sl KVM} and {\sl VirtualBox} VMs.

Figure~\ref{fig:mastercodecompile} shows the average compilation time of {\tt MasterCode} over ten repetitions of the benchmark normalized to the native host result.
The standard mean deviation is shown in black on the top of each bar. The baseline host result is roughly $\sim 923$ seconds.
We notice that the first execution of the test is always the slowest, for all environments. This is due to I/O caching effects that disappear from the second iteration onward,
where on the other hand we find a good stability. Indeed the standard deviation is small in all cases.
From the results, we observe that {\sl udocker 1.0} and {\sl udocker 1.0.2} take a performance hit of about $30\%$ with respect to the native host. Both these versions were
run using the standard PROOT model without SECCOMP filtering. On the other hand, the newest udocker release, used with the SECCOMP filtering enabled, is only about $5\%$ slower than
the host and very close to the timing obtained in a {\sl KVM} environment. The {\sl VirtualBox} environment has a compilation time between the {\sl KVM} and
the old versions of {\sl udocker}, with a performance hit of about $15\%$. The {\sl Docker} environment is the very close to the native performance.

The results of the sampling-benchmark are shown in Figure~\ref{fig:mastercodecmssm}, with the same format used for the compilation case.
We performed the same sampling three times to avoid caching biases and statistical oscillations due to other background processes. The average running time for the native host
is of $\sim 31187$ seconds. We observe that all the different setups yield running times that are at most about $2.5\%$ slower than the host machine.
In this case, the slowest environment is the {\sl KVM} one, while {\sl Docker} is the fastest, with {\sl udocker} (P2 mode) and {\sl KVM} following very closely.
The striking difference in the observed pattern with respect to the compilation--test case, is due to the fact that the sampling--phase is mostly CPU-bound, while the compilation--phase is mostly
dependent on the I/O performances of the environment/system. Indeed, {\sl udocker}, even in its default PROOT mode (without SECCOMP filtering), shows a negligible performance hit with respect
to the native host.

\section{MPI parallel execution: OpenQCD}

In this section we describe how to use {\sl udocker} to run MPI jobs in HPC systems. The process described applies both to interactive and batch--style job submission, following the philosophy of {\sl udocker}; it does not require any special user privileges, nor system administration intervention to setup.

We have chosen openQCD~\cite{OPENQCD} to investigate both the applicability of our solutions and the effects of {\it containerizing} MPI processes on the performance of the code. The reason to choose this application stems from its high impact and the code features.

Regarding impact, Lattice QCD simulations spend every year hundreds of millions of CPU hours in HPC centers in Europe, USA and Japan. Current competitive simulations spread over thousands of processor cores. From the point of view of parallelization it requires only communication to nearest neighbors and thus, presents nice properties regarding scalability. OpenQCD is one of the most optimized codes available to run Lattice simulations and therefore, is widely distributed.

It is implemented using only open source software, may be downloaded and used under the license terms of the {\it GPL} model. Therefore it is an excellent laboratory for us to investigate with the required clarity the performance and applicability of our solution.

Competitive QCD simulations require latencies on the order of one microsecond, and Intranet bandwidths on the range 10--20 Gb/second in terms of effective throughput. This is currently achieved only by the most modern Infiniband fabric interconnects (from QDR on). In what follows, we describe the steps to execute openQCD using {\sl udocker} in a HPC system using Infiniband as low--latency interconnect. An analogous procedure can be followed for other MPI applications, also via simpler TCP/IP interconnects.

A container image of openQCD can be downloaded from the public {\it Docker Hub} repository\footnote{https://hub.docker.com/r/iscampos/openqcd}. From this image a container can be extracted to the filesystem with {\tt udocker create}, as described before.

In the {\sl udocker} approach the {\tt mpiexec} of the Host machine is used to submit $np$ MPI process instances as containers, in such a way that the containers are able to communicate via the Intranet fabric network (Infiniband in the case at hand).

For this approach to work, the code in the container needs to be compiled with the same version of MPI that is available in the HPC system. This is necessary because the Open MPI versions of {\tt mpiexec} and {\tt orted} available in the host system need to match with the compiled program. This limitation has to do with incompatibilities of the Open MPI libraries across different versions as well as with tight integration of Open MPI with the batch system.

The MPI job submission to the HPC cluster succeeds by issuing the following command:

\newpage

\begin{verbatim}
$HOST_OPENMPI_BIN/mpiexec -np 128 udocker run \
                          --hostenv --hostauth --user=$USERID \
                          --workdir=$OPENQCD_CONTAINER_DIR \
                          openqcd \
                          $OPENQCD_CONTAINER_DIR/ym1 -i ym1.in
\end{verbatim}

Where the environment variable {\tt HOST\_OPENMPI\_BIN} points to the location of the host MPI installation directory

Depending on the application and host operating system, a variable performance degradation may occur when using the default execution mode ({\it Pn}). In this situation other execution modes (such as {\it Fn}) may provide a significant higher performance

\subsection{Scaling tests}

\begin{figure}
  \centering
  \includegraphics[width=\textwidth]{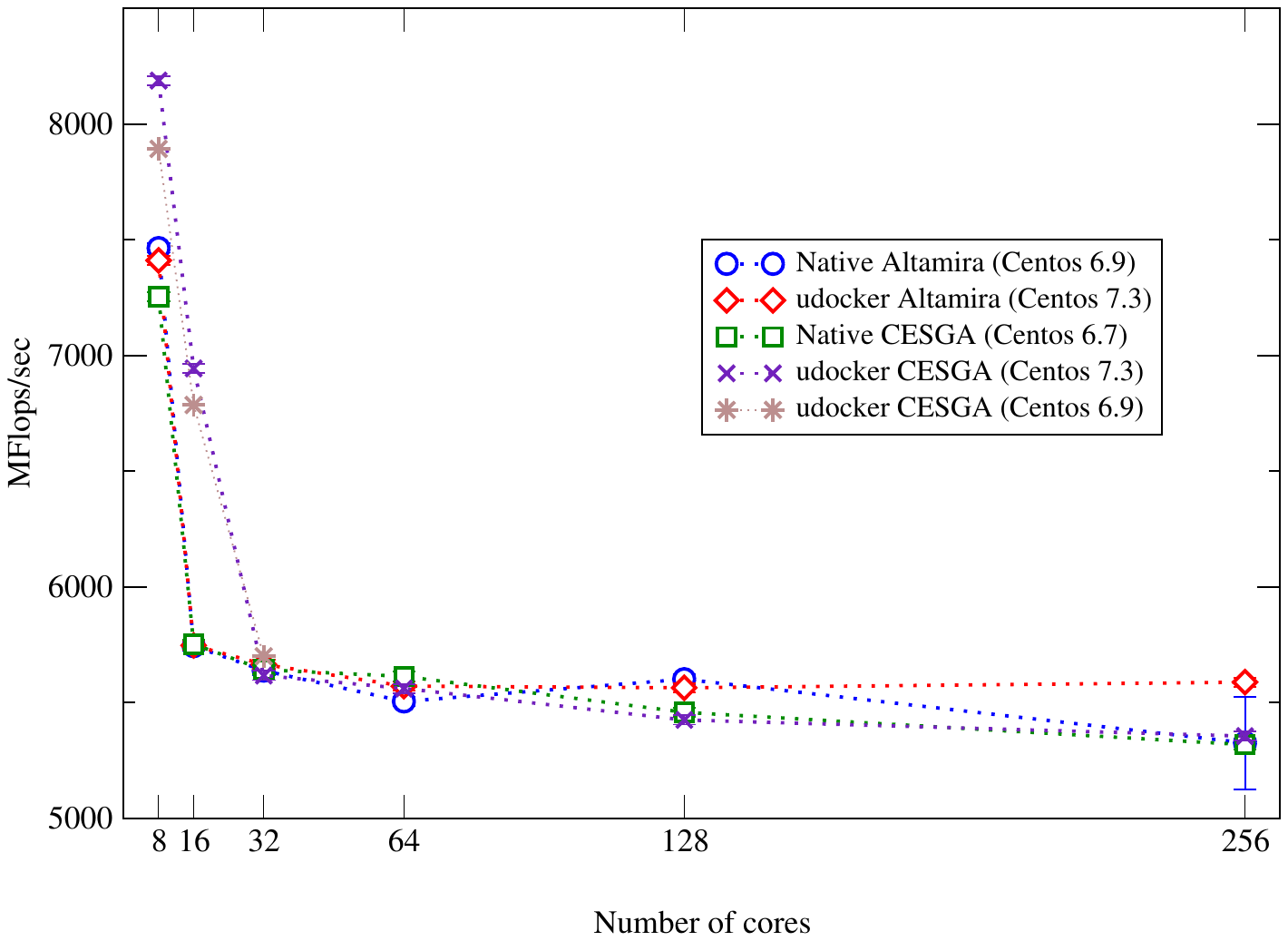}
  \caption{Weak scaling analysis of the performance of the Lattice Dirac Operator (local lattice $32^4$ ) using a containerized version of the application with {\sl udocker}, compared with the performance of the code used in the classical way (native performance). In these tests we have used {\tt P1} as execution mode. The results do not vary using {\tt F1}.}
  \label{fig:scalingQCD}
\end{figure}

In Figure \ref{fig:scalingQCD} we plot the (weak) scaling performance of the most common operation in QCD, that is, applying the so called Dirac Operator to a spinor field \cite{LECTURESQCD}. This operation can be seen as a sparse matrix--vector multiplication across the whole Lattice, of a matrix of dimension proportional to the Volume of the Lattice ($12\times T\times L^{3}$) times a vector of the same length.

The scaling properties have been measured in two different HPC systems: Altamira and CESGA\footnote{See Appendix for a hardware/software description of both machines.}.

As can be observed the performance using {\sl udocker} to run the MPI jobs is at the very least equal to the native performance. There are however a couple of observations to be made:

\begin{itemize}
\item At CESGA we observe that the performance of udocker is substantially better than the native performance when using 8 and 16 MPI processes, i.e., when the MPI processes spans only within one single node (the nodes at CESGA have 24 cores).

The host machine is running a CentOS 6.7 OS, while the container has a CentOS 7.3. We have checked, by running also the application with udocker in a CentOS 6.9 container, that this improvement is due to the improvement in the libraries of newer versions of CentOS. We actually observe that the performance increases smoothly when increasing the version of CentOS.

More advanced versions of the libraries in the Operating System are able to make a more efficient usage of the shared memory capabilities of the CESGA nodes. We do not observe such improvement when the run is distributed over more than one node, in this case the effects of the intra-network communication overheads are likely dominating the performance.

It is interesting to observe how the usage of containers can improve the applications performance when the underlying system has a somewhat older Operating System. This fact is specially relevant in most HPC centers that, for stability reasons, tend to have a very conservative policy when it comes to upgrading the underlying operating system and software.

\item In Altamira we observe a large dispersion in the performance results using 256 cores in native mode. The performance is much more stable using {\sl udocker}. This can be traced down to local GPFS filesystem issues, which are more visible in the native run, as the system needs to look for the openMPI libraries across the filesystem, while with {\sl udocker} the container has everything included in the {\tt \$UDOCKER\_DIR} directory.
\end{itemize}

Besides the scaling tests we also run openQCD for various settings of Lattice size and parameters, encountering no impact in performance in the time employed for each Hybrid MonteCarlo trajectory. This is of no surprise as I/O is not particularly intensive and frequent in Lattice QCD simulations: the usual checkpoints generate only Kbytes of data, while configurations, which in our Lattice sizes reached up to 1GB in size, are written in intervals of a few hours. For such I/O rates using {\sl udocker} implies no penalty in performance.

Related to the above is the fact that in all the tests performed we have observed no performance difference between using {\tt P1} or {\tt F1} execution modes.

In summary, exploiting low-latency computing facilities using containers technology appears as an appealing possibility to maximize the efficiency on the use of resources, while guaranteeing researchers autonomy in profiting from modern system software capabilities.

\section{Accessing GPGPUs: Modeling of Biomolecular Complexes}

DisVis \cite{DISVIS} and PowerFit \cite{POWERFIT} are two software packages for modeling biomolecular complexes. A detailed description of both software packages can be found in \cite{HADDOCK}.

These software packages are coded in Python and are open source under the MIT license, published in github \cite{GITHADDOCK}. They leverage the use of GPGPUs through the OpenCL framework (PyOpenCL Python package) since the modeling uses the evaluation of Fast Fourier Transforms (FFTs), it uses the packages clFFT\footnote{https://github.com/clMathLibraries/clFFT} and gpyfft\footnote{https://github.com/geggo/gpyfft}.

Gromacs \cite{GROMACS} is a versatile package to perform molecular dynamics, i.e. simulate the Newtonian equations of motion for systems with hundreds to millions of particles. It is primarily designed for biochemical molecules like proteins, lipids and nucleic acids that have a complex bonded interactions, but since Gromacs is extremely fast at calculating the non--bonded interactions (that usually dominate simulations) many researchers are also using it for non--biological systems, such as polymers.

\subsection{Setup and deployment}

The physical host used to benchmark {\sl udocker} with the DisVis and Gromacs applications using GPGPUs was made available by the Portuguese National Distributed Computing Infrastructure (INCD). It is hosted at the INCD main datacenter in Lisbon and its characteristics are described in the Appendix.

The applications DisVis v2.0.0 and Gromacs 2016.3 were used. Docker images were built from Dockerfiles for both applications and for two operating systems:  CentOS 7.3 (Python 2.7.5, gcc 4.8.5) and Ubuntu 16.04 (Python 2.7.12, gcc 5.4.0), the versions of the applications in the Docker images are the same as in the physical host.

The applications in the Docker images were built with GPGPU support and the NVIDIA software that matches the same version of the NVIDIA driver deployed in the physical host system.

Those same images were then used to perform the benchmarks using both Docker and {\sl udocker}. In particular, {\sl udocker} has several modes of execution (c.f. the user manual hosted at \cite{UDOCKER}), available through the {\tt setup --execmode} option. Two such options have been chosen for Gromacs execution, namely: {\tt P1} corresponding to PRoot with SECCOMP filtering and {\tt F3} corresponding to Fakechroot with maximum isolation from the host system. For DisVis we only used the {\tt P1} execution mode.

Table~\ref{table:res_gpus} summarizes the different execution environments used in the benchmarks in terms of operating system and type of machine: physical (the host), Docker or {\sl udocker}. The tags UDockP1 and UDockF3 correspond to the two chosen execution modes of {\sl udocker} mentioned above.

\begin{table}
  \begin{center}
    \begin{tabular}{l|c|c} \toprule
    \textbf{tag} & OS & type of machine \\
    \hline
    \textbf{Phys-C7-QK5200} & CentOS 7 & Physical \\
    \textbf{Dock-C7-QK5200} & CentOS 7 & Docker \\
    \textbf{Dock-U16-QK5200} & Ubuntu 16 & Docker \\
    \textbf{UDockP1-C7-QK5200} & CentOS 7 & udocker mode P1\\
    \textbf{UDockP1-U16-QK5200} & Ubuntu 16 & udocker mode P1\\
    \textbf{UDockF3-C7-QK5200} & CentOS 7 & udocker mode F3\\
    \textbf{UDockF3-U16-QK5200} & Ubuntu 16 & udocker mode F3\\
    \end{tabular}
  \end{center}
  \caption{Execution environments for the DisVis and Gromacs applications.}
  \label{table:res_gpus}
\end{table}

\subsection{Benchmark results}

The DisVis use case was executed with input molecule {\tt PRE5-PUP2 complex}. The application is executed with one GPU (the option {\tt -g} selects the GPU execution mode), using the following command:

\begin{verbatim}
$shell> disvis O14250.pdb \
               Q9UT97.pdb \
               restraints.dat \
               --angle 5.0 \
               --voxelspacing 1 \
               -g \
               -d ${OUT_DIR}
\end{verbatim}

The Gromacs {\tt gmx mdrun} was executed with an input file of the amyloid beta peptide molecule. The application is executed with one GPU and 8 OpenMP threads (per MPI rank), using the following command:

\begin{verbatim}
$shell> gmx mdrun -s md.tpr \
                  -ntomp 8 \
                  -gpu_id 0
\end{verbatim}

\begin{figure}
  \centering
  \includegraphics[width=0.95\textwidth]{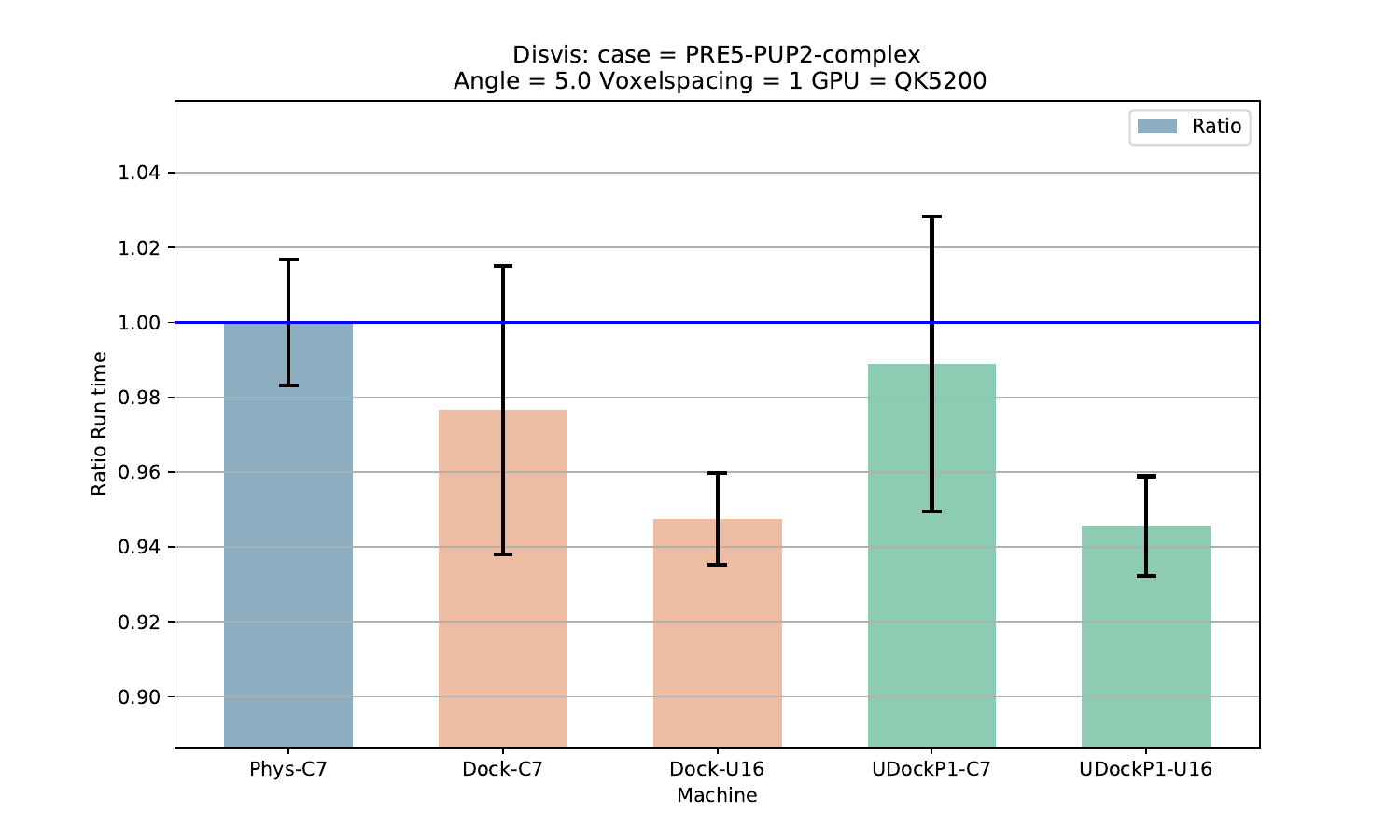}
  \caption{Ratio between runtime values of Docker or {\sl udocker} and the physical host of the DisVis use case. Lower values indicate better performance.}
  \label{fig:ratiodisvis}
\end{figure}

\begin{figure}
  \centering
  \includegraphics[width=0.95\textwidth]{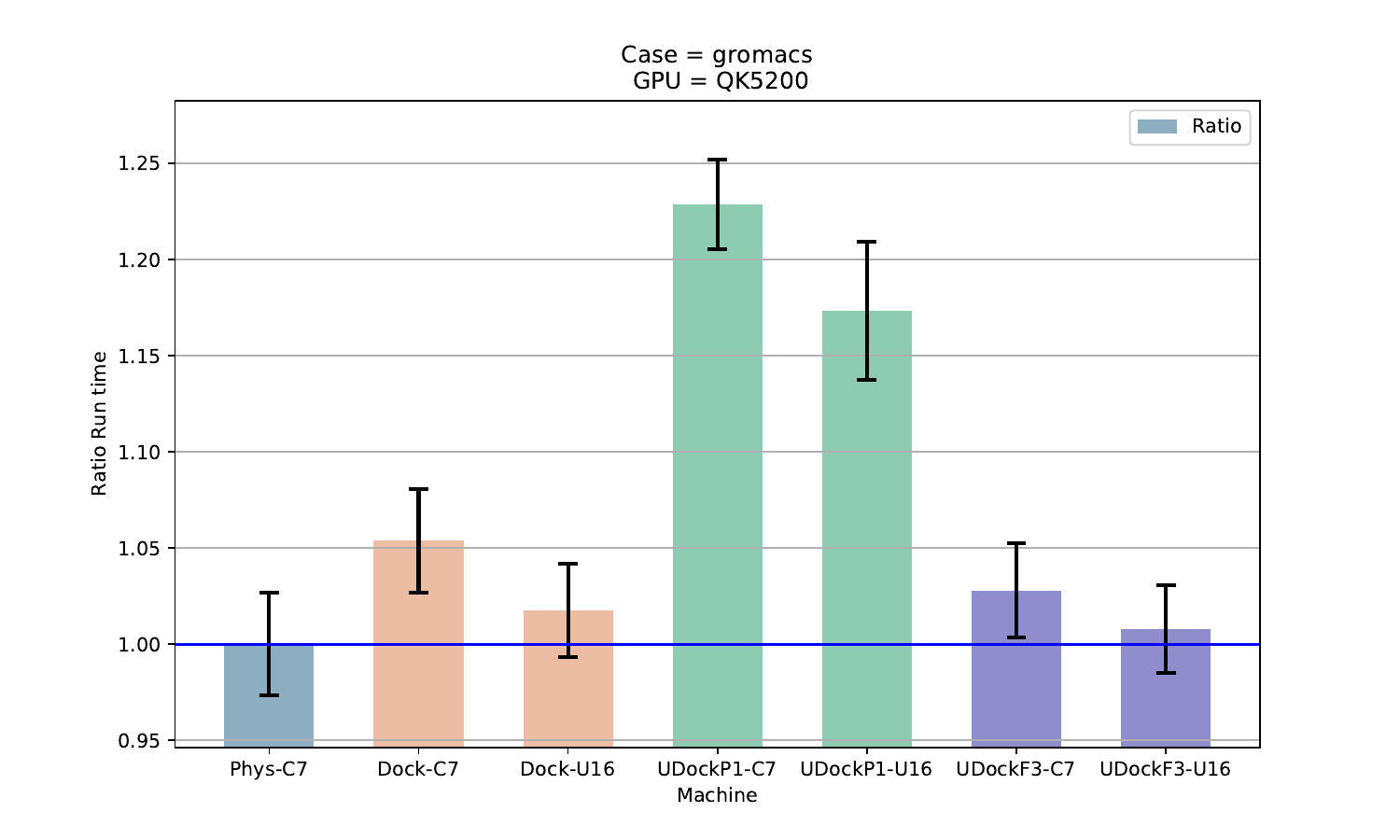}
  \caption{Ratio between runtime values of Docker or {\sl udocker} and the physical machine of the Gromacs use case. Lower values indicate better performance.}
  \label{fig:ratiogromacs}
\end{figure}

For each case, 20 runs have been executed for statistical purposes. The run time of each execution is recorded and a statistical analysis is done as follows: The outliers are detected and masked from the sample of 20 points, then the average and standard deviation are calculated for each case and environment option (physical host, Docker or {\sl udocker}). The ratio between the average run time of each docker or {\sl udocker} environment with the average run time in the physical host is plotted in Figures~\ref{fig:ratiodisvis} and \ref{fig:ratiogromacs}. Typical execution times for both cases range between 20 and 30 minutes.

The first ratio (column) in each figure is 1 since the baseline is the physical host, the standard deviation of the ratio is calculated based on the statistical formula for the ratio of two variables:

\begin{equation}
\Delta R = t_i/t_{host} \times \sqrt{ (\Delta t_i / t_i)^2 + (\Delta t_{host} / t_{host})^2 }
\end{equation}

Where: $t_i$ is the average runtime in a given environment, $t_{host}$ is the average runtime in the physical host, $\Delta t_{host}$ and $\Delta t_i$ are the corresponded standard deviations.

Figure \ref{fig:ratiodisvis} correspond to the benchmark of the DisVis use case. It can be seen that the execution of both Docker and {\sl udocker} in CentOS 7.3 containers although $1-2\%$ smaller than 1 is still compatible with this value. This shows that executing the application in this type of virtualized environment has the same performance as executing it in the physical host. On the other hand executing the application in the Ubuntu 16.04 containers (both Docker and {\sl udocker} cases) the execution time is around 5$\%$ better than in the physical host.

Figure \ref{fig:ratiogromacs} correspond to the benchmark of the Gromacs use case. It can be seen that the execution in CentOS 7.3 containers for Docker (second bar) and {\sl udocker} with execution mode {\tt F3} (Fakechroot with maximum isolation from the host system, sixth bar), are 3--5$\%$ worse than the execution in the physical host although still compatible with one due to the statistical error. Also in the case of execution in Ubuntu 16.04 containers (third and seventh bars) the ratios are compatible with one, i.e. the containerized application execution has the same performance as in the physical host. In the case of executing the application in {\sl udocker} containers with execution mode {\tt P1} (PRoot with SECCOMP filtering, forth and fifth bars), a penalty in performance of up to 22$\%$ is observed.

The difference in performance between {\sl udocker} execution modes {\tt F3} and {\tt P1} for the Gromacs use case can be explained by the computational model itself, since Gromacs is using the GPGPU and 8 OpenMP threads (corresponding to 8 CPU cores) frequent exchanges of data between the GPGPU and the CPU threads and possible context change between CPU tasks translate into a performance penalty in the {\tt P1} execution mode that does not occur in the
{\tt F3} execution mode. This is partially confirmed by comparing to the DisVis use case that uses almost exclusively the GPGPU (and at most only 1 CPU) where the {\tt P1} execution mode does not have any impact on performance.

\section{Conclusions}

In this paper we have described a middleware suite that provides full autonomy to the user when it comes to execute applications in Docker container format. We have shown that our solution provides a complete encapsulation of the software, without significantly impacting performance.

In this paper we have analyzed a number of differently oriented use cases, and got clear evidence that compute intensive applications pay practically no performance penalty when executed as {\sl udocker} containers.
Regarding I/O bounded applications, the {\sl udocker} execution modes {\sl Pn} introduce a penalty that can be largely minimized by using the {\sl Fn} execution modes. The small penalty still observed with the {\sl Fn} modes is introduced by the filename translation, the data access itself is unaffected. Other tools such as Singularity or Shifter can mount container images as file systems keeping many file system calls local. Mounting the images may yield better performance especially when using a shared network file system, however this has the disadvantage of requiring privileges. All OpenQCD performance measurements in this paper were performed with the container image placed in Lustre file systems without any noticeable impact in comparison with native execution.

Tools like {\sl udocker} are essential when it comes to access external computing resources with different execution environments. It is particularly so when accessing cloud infrastructures and shared computing infrastructures such as HPC clusters, where running software with encapsulation is a necessity to remain independent of the host computing environment. Users of DisVis and PowerFit have been running these applications in the EGI Grid infrastructure at sites offering GPGPU resources, using {\sl udocker} and performing grid job submission as any other application. {\sl udocker} empowers users to execute applications encapsulated in Docker containers in such systems in a fully autonomous way.

Since its first release in June 2016 {\sl udocker} expanded quickly in the open source community as can be seen from the github metrics (number of stars, forks or number of external contributions). Since then it has been used in large international collaborations like the case of MasterCode reported here.

The extra flexibility offered by {\sl udocker} has made it a very appreciated tool, and has been already adopted by a number software projects to complement Docker.
Among them {\sl openmole}\footnote{https://www.openmole.org/api/index.html\#org.openmole.plugin.task.udocker.package}, {\sl bioconda}\footnote{https://anaconda.org/bioconda/udocker},
{\sl common-workflow-language (cwl)}\cite{CWL} or {\sl SCAR - Serverless Container-aware ARchitectures}\footnote{https://github.com/grycap/scar}.

\section*{Acknowledgements}
This work has been performed in the framework of the H2020 project INDIGO-Datacloud (RIA 653549).
The work of E.B.~is supported by the Collaborative Research Center SFB676 of the DFG, ``Particles, Strings and the early Universe''.
The proofs of concept presented have been performed at the FinisTerrae II machine provided by CESGA (funded by Xunta de Galicia and MINECO), at the Altamira machine (funded by the University of Cantabria and MINECO) and at the INCD-{\it Infraestrutura Nacional de Computa\c{c}\~ao Distribu\'ida} (funded by FCT, P2020, Lisboa2020, COMPETE and FEDER under the project number 22153-01/SAICT/2016).

We are indebted to the managers of these infrastructures for their generous support and constant encouragement.

\newpage

\section*{Appendix: Hardware and Software setup}

The images of the containers used in this work are publicly available in the Docker hub, at the following urls:

\begin{itemize}
\item {\tt OpenQCD}: https://hub.docker.com/r/iscampos/openqcd/
\item {\tt DISVIS}: https://hub.docker.com/r/indigodatacloudapps/disvis
\item {\tt POWERFIT}: https://hub.docker.com/r/indigodatacloudapps/powerfit
\item {\tt MASTERCODE}: https://hub.docker.com/r/indigodatacloud/docker-mastercode/
\end{itemize}

\noindent
The proofs of concept presented have been carried out in the following infrastructures, with the following Hardware/OS/Software setup:

\begin{itemize}

\item{Finisterrae-II - CESGA, Santiago de Compostela}

\begin{verbatim}https://www.cesga.es/es/infraestructuras/computacion/FinisTerrae2
\end{verbatim}

Hardware Setup:

\begin{itemize}
\item {Processor type} Intel(R) Xeon(R) CPU E5-2680 v3 @ 2.50GHz, with cache size 30MB
\item {Node configuration} 2 CPUs Haswell 2680v3, (2x12 = 24 cores/node) and 128GB RAM/node
\item {Infiniband Network} Mellanox Infiniband FDR@56Gbps
\end{itemize}

Software setup:
\begin{itemize}
\item{Operating System} CentOS 6
\item{Compilers}
GCC 6.3.0
\item{MPI libraries}
openmpi/2.0.2
\end{itemize}

\item{Altamira - IFCA-CSIC, Santander}

\begin{verbatim}
https://grid.ifca.es/wiki/Supercomputing/Userguide
\end{verbatim}

Hardware Setup:
\begin{itemize}
\item {Processor type} Intel(R) Xeon(R) CPU E5-2670 0 @ 2.60GHz, with cache size 20MB.
\item {Node configuration} 2 CPUs E5-2670, (2x8 = 16 cores/node) and 64GB RAM/node
\item {Infiniband Network} Mellanox Infiniband FDR@56Gbps
\end{itemize}

Operating system and Software setup:
\begin{itemize}
\item{Operating System} CentOS 6
\item{Compilers}
GCC 5.3.0
\item{MPI libraries}
openMPI/1.8.3
\end{itemize}

\item{INCD, Lisbon}

\begin{verbatim}
https://www.incd.pt
\end{verbatim}

Hardware Setup:
\begin{itemize}
\item {Processor type} Intel(R) Xeon(R) CPU E5-2650L v3 @ 1.80GHz
\item {Node configuration} 2 CPUs (2x12 = 24 cores/node x2 = 48 in HiperThreading mode) and 192GB RAM
\item {GPU model} NVIDIA Quadro K5200
\end{itemize}

Operating system and Software setup:
\begin{itemize}
\item{Operating System} CentOS 7.3
\item{NVIDIA driver} 375.26
\item{CUDA} 8.0.44
\item{docker-engine} 1.9.0
\item{udocker} 1.0.3
\end{itemize}

\end{itemize}

\end{document}